# Character of the Dislocation Bands in the (A+B) regime of the Portevin-Le Chatelier effect in Al-2.5%Mg alloy

A. Chatterjee, A. Sarkar, P. Barat*, P. Mukherjee and N. Gayathri

Variable Energy Cyclotron Centre

1/AF Bidhan Nagar, Kolkata 700064, India

## Abstract

*The Portevin-Le Chatelier(PLC) effect has been investigated by deforming Al-2.5%Mg alloy in the strain rate regime where simultaneously two types (type B and type A) of serrations appear in the stress strain curve. Our analysis reveal that in this strain rate regime the entire PLC dynamics for a particular strain rate experiment is governed by a single band which changes its character during the deformation.*

The Portevin-Le Chatelier (PLC) effect denotes the jerky flow or the repetitive yielding observed in the stress-strain curve of many dilute interstitial and substitutional metallic alloys deformed in a certain range of strain rates and temperatures [1-5]. This phenomenon is associated with the loss of ductility and surface roughness of these materials. The serrations in the stress-strain curve are found to have one to one relationship with the surface markings or band on the specimen, which indicates that each stress drop is associated with the nucleation of a deformation band at some point in the specimen. This band propagates along the tensile axis of the specimen. With increasing strain rate the bands show an increasing tendency to propagate, while the observed stress serrations are found to decrease in magnitude. These bands are well characterized in polycrystals and are classified into three generic types [6-8]. At low strain rates randomly nucleated static type C bands are observed which are identified with large characteristic load drops on the stress-strain curve. At medium strain rates type B, hopping bands are seen. The serrations formed owing to its movement are regular with amplitude smaller than that of type C. In contrast, at high strain rates the continuously propagating type A bands associated with small stress drops are observed.

* Corresponding author: pbarat@veccal.ernet.in

Confusion exists among the researchers regarding the exact cause of the PLC effect. However, a general consensus is that the origin of this plastic instability is the dynamic interaction between the defect populations, namely the mobile dislocations and the solute atoms [9-11].

Due to the inherent nonlinearity and the presence of multiple timescales, nonlinear dynamics has usually found its way in the study of PLC effect. In the last two decades many statistical and dynamical analysis have been carried out on the PLC effect [12-16]. These studies provided a reasonable understanding of the effect. Dynamical analysis suggested that distinct features could be associated with each type of band observed in the PLC effect [12-18]. In the case of type B bands a chaotic dynamics was demonstrated [17]. For high strain rates i.e. for type A band regime, the dynamics exhibits self organized critical (SOC) nature [15].

Al-Mg alloy with nominal percentage of Mg shows PLC effect at room temperature for a wide range of strain rates. All the three types of serrations are observed in the stress-strain curve of this alloy system. Consequently the Al-Mg alloy has become a model system to study the dynamical features of the PLC effect. In case of Al-2.5% Mg alloy at some intermediate strain rate regime simultaneous occurrence of two types (type B and type A) of serrations was reported recently [19]. However, no work has been reported so far to address whether the two different types of bands coexist or a single band changes its character during the deformation.

Tensile tests were conducted on flat specimens prepared from polycrystalline Al-2.5%Mg alloy over a wide range of strain rate from $8.02\times10^{-5}s^{-1}$ to $2.02\times10^{-3}s^{-1}$. Specimens with gauge length, width and thickness of 25, 5 and 2.3 mm, respectively were tested at room temperature (300K) in an INSTRON (model 4482) machine. The stress-time response was recorded at a sampling rate of 20 Hz. In this paper, we are interested to study the PLC effect in the strain rate regime where both type B and type A band character is observed. Taking clue from the work reported in the reference 19 and from the statistical analysis presented subsequently in this paper, we can identify the (A+B) regime for Al-2.5%Mg alloy from the strain rate $2\times10^{-4}s^{-1}$ to $9\times10^{-4}s^{-1}$. In addition to the representative data from this regime (A+B), one data each from Type A (strain rate $9.92\times10^{-4}s^{-1}$ ) band regime and Type B (strain rate $8.02\times10^{-5}s^{-1}$) band regime are analyzed and presented here. Fig. 1 shows the observed PLC effect in a typical stress-strain curve in the (A+B) regime at the strain rate $3.89\times10^{-4}$ $S^{-1}$. The inset shows the magnified view of stress-time variation of a typical region in it.

A quantitative study of the distribution functions of the stress drop magnitudes obtained from distinct strain rate experiments and their consequent classification is a fundamental step towards a better understanding of the underlying dynamics of the PLC effect. Lebyodkin et al. [12-14] pioneered this statistical analysis and in Al-Mg alloy they could identify that in the intermediate strain rate regime, i.e. for type B serrations, the stress drop magnitudes follow a peaked distribution and for type A serrations, the stress drop magnitude shows a power law behavior. Based on this idea, we first concentrated on the study of the statistics of the magnitudes of the stress drops of the PLC effect observed at the strain rate regime where the presence of both type B and type A serrations are reported [19]. We have calculated the frequency distribution of the stress drop magnitudes obtained from the stress time series data recorded during the PLC effect. Fig. 2 shows the frequency distribution of the stress drop magnitudes for the experiments conducted at different strain rates on the Al-2.5%Mg alloy. The stress drop magnitude distribution obtained from the lowest strain rate shows a peaked distribution. The peaked distribution is associated with the type B serrations [12-14]. Subsequent plots in Fig. 2 show two distinct behavior. Smaller stress drop values follow a power law type distribution and the comparatively larger stress drop values show a peaked distribution. The power law distribution of the stress drop magnitudes is reminiscent to the type A serrations [12-14]. The simultaneous appearance of these two distributions in Fig. 2, supports the existence of both type B and type A serrations during the deformation in the mentioned strain rate regime. With increase in strain rate, the power law type distribution becomes predominant over the peaked one. At sufficiently high strain rate of $9.92\times10^{-4}\ s^{-1}$, only the power law distribution persists which indicates the transition to type A band region. This transition of the system from type B band to type A band regime is also revealed from the distribution of number of stress drops per unit time interval. Type B band region is characterized by fewer number of stress drops per unit time whereas type A band region gives more frequent stress drops [20]. These are reflected in the fact that the occurrence frequency of lesser number of stress drops per unit time is higher for type B band and the opposite scenario is exhibited in case of type A band. Fig. 3 shows the distribution of number of stress drops per unit time at different strain rates. The distribution resembles that of type B and type A band at the lowest and the highest strain rates respectively. But in the intermediate region it shows a peaked distribution.

From Fig. 3, we can identify the transition region from type B to type A band regime with strain rate. We are mainly interested in this transition region and further analysis are done to investigate whether the two types of bands exist simultaneously.

Fig. 2 supports the fact that in the strain rate regime $3.89\times10^{-4}$ $S^{-1}$ to $6.57\times10^{-4}$ $S^{-1}$ in Al-2.5%Mg alloy both type B and type A serrations are present in the fixed strain rate experiments. In recent years, many sophisticated experimental techniques have been employed to investigate the kinematics of the PLC bands [21-24] by observing its movement in the sample. All these studies are confined to a particular type of the PLC band, either type A or type B. Nevertheless, no attempt has been made so far to visualize the movement of the bands where both types exist. One fundamental question of interest arises in this context is, whether the dynamics of the PLC effect in this intermediate strain rate regime is governed by two bands (type B and type A) together or a single band changing its character from type B to type A or vice versa during its movement. The answer to this basic and interesting question is not straightforward. We anticipate that the available visual techniques used to study the dynamics of the PLC band, will not be efficient enough to distinguish between the type B and type A bands in the sample. Hence, in order to get a certain conclusive answer to this important question, we have adopted a unique statistical approach to investigate the stress time series data recorded during the PLC effect. We have calculated the probability of occurrences of the stress drops in the different chronological order e.g. AA, BB, AAA, AB,…, etc.. Here, AA indicates the occurrence of two consecutive type A serrations. The plots of the probability of occurrences of different combination of successive serrations for different strain rates are shown in Fig. 4. It is seen in Fig. 4 that the probabilities of occurrence of BB, AA, BBB and AAA i.e. consecutive similar type of serrations are high whereas the probabilities of occurrence of AB, BA i.e. two different types of serrations are low. High occurrence probability of consecutive similar type of serrations suggests the existence of only one band in the sample.

The nature of the PLC band is primarily governed by the number of dislocations participating in the formation of the band. At high strain rate, the strong internal stresses associated with the dislocations, give rise to a highly correlated movement of only a few number of dislocations resulting in a narrow band. Thus, it can move continuously under such high pulling rate. Even though new dislocations are always generated, this strong correlation restricts the band width and the band continues to be of type A. On the other hand, in the strain rate regime where type B bands are observed, the strain rate is not sufficient enough to provide such strong correlation among the dislocations. Instead, a large number of dislocations correlate weakly forming a wider band which carries the signature of type B. In the intermediate region of strain rate where both type B and type A bands are observed, the correlation among the dislocations is moderate. This can allow the changes in the number of dislocations in the band and in turn changes its character from type B to type A and vice versa.

The general distinction between type B and type A bands in the PLC effect is by the number of dislocation participating in it. However, it is rather impossible to define a critical number of dislocations in the band which marks the distinction between type B and type A band.

In conclusion, our analysis of the stress time series data recorded during the PLC effect reveals that in the intermediate strain rate regime the dynamics of the PLC effect is governed by a single dislocation band which changes its character during its movement.

## References

[1] P.G. McCormick, Acta Metall. 20, 351 (1972).

[2] K. G. Samuel, S. L. Mannan and P. Rodriguez, Acta Metal. 36, 2323 (1988).

[3] E. Pink, Scripta Metall et Mater. 30, 767 (1994).

[4] G. G. Saha, P. G. McCormick and P. Rama Rao, Mater. Sci. & Eng. 62, 187 (1984).

[5] A. Sarkar, A. Chatterjee, P. Barat, P. Mukherjee, Mater. Sci. & Eng. A, 459, 361 (2007).

[6] K. Chihab, Y. Estrin, L.P. Kubin, J. Vergnol, Scripta Metall. 21, 203 (1987).

[7] E. Rizzi, P. Hahner, Int. J. Plasticity 20, 121(2004).

[8] L.P. Kubin, C. Fressengeas, G. Ananthakrishna, Dislocations in Solids, Vol. 11, ed. F. R. N. Nabarro, M. S. Duesbery (Elsevier Science, Amsterdam, 2002, p 101).

[9] A.H. Cottrell, Dislocations and plastic flow in crystals (Oxford University Press, London, 1953).

[10] A. Van den Beukel, Physica Status Solidi(a) 30, 197(1975).

[11] J. Schlipf, Scripta Metall. Mater. 31, 909 (1994).

[12] M.A. Lebyodkin, Y. Brechet, Y. Estrin, L.P. Kubin, Phys. Rev. Lett. 74, 4758 (1995).

[13] M. Lebyodkin, Y. Brechet, Y. Estrin, L.P. Kubin, Acta Mater. 44, 4531(1996).

[14] M. Lebyodkin, L. Dunin-Barkowskii, Y. Brechet, Y. Estrin, L.P. Kubin, Acta Mater. 48, 2529(2000).

[15] G. Ananthakrishna, S. J. Noronha, C. Fressengeas and L. P. Kubin, Phys. Rev. E 60, 5455 (1999).

[16] M. S. Bharathi, M. Lebyodkin, G. Ananthakrishna, C. Fressengeas and L. P. Kubin, Acta Mater. 50, 2813 (2002).

[17] G. Ananthakrishna, C. Fressengeas, M. Grosbras, J. Vergnol, C. Engelke, J. Plessing, H. Neuhauser, E. Bouchaud, J. Planes, L.P. Kubin, Scr. Metall. Mater. 32, 1731 (1995).

[18] M.S. Bharathi, M. Lebyodkin, G. Ananthakrishna, C. Fressengeas, L.P. Kubin, Phys. Rev. Lett. 87, 165508 (2001).

[19] K. Chihab, C. Fressengeas, Mat. Sci. Eng. A 356, 102 (2003).

[20] M. Lebyodkin, C. Fressengeas, G. Ananthakrishna and L. P. Kubin, Mat. Sci. Eng. A 319 (2001), 170.

[21] R. Shabadi, S. Kumar, H. J. Roven, E. S. Dwarakadasa, Mat. Sci. Eng. A 364, 140 (2004).

[22] N. Ranc and D. Wagner, Mater. Sci. and Eng. A 394, 87 (2005).

[23] Q. Zhang, Z. Jiang, H. Jiang, Z. Chen, X. Wu, International Journal of Plasticity 21, 2150 (2005).

[24] P. Hähner, A. Ziegenbein, E. Rizzi, and H. Neuhäuser, Phys. Rev. B 65, 134109 (2002)

## Figure Captions:

Fig. 1 True Stress vs. True Strain curve of Al-2.5%Mg alloy at a strain rate of $3.89\times10^{-4}$ $Sec^{-1}$. The inset shows a typical region of the curve in the Stress-Time plot.

Fig. 2 Frequency distribution of the stress drop magnitudes for experiments conducted at different strain rates on Al-2.5%Mg alloy.

Fig. 3 Distribution of number of stress drops per unit time at different strain rates.

Fig. 4 Probability of occurrences of different combination of successive serrations for different strain rate experiments.

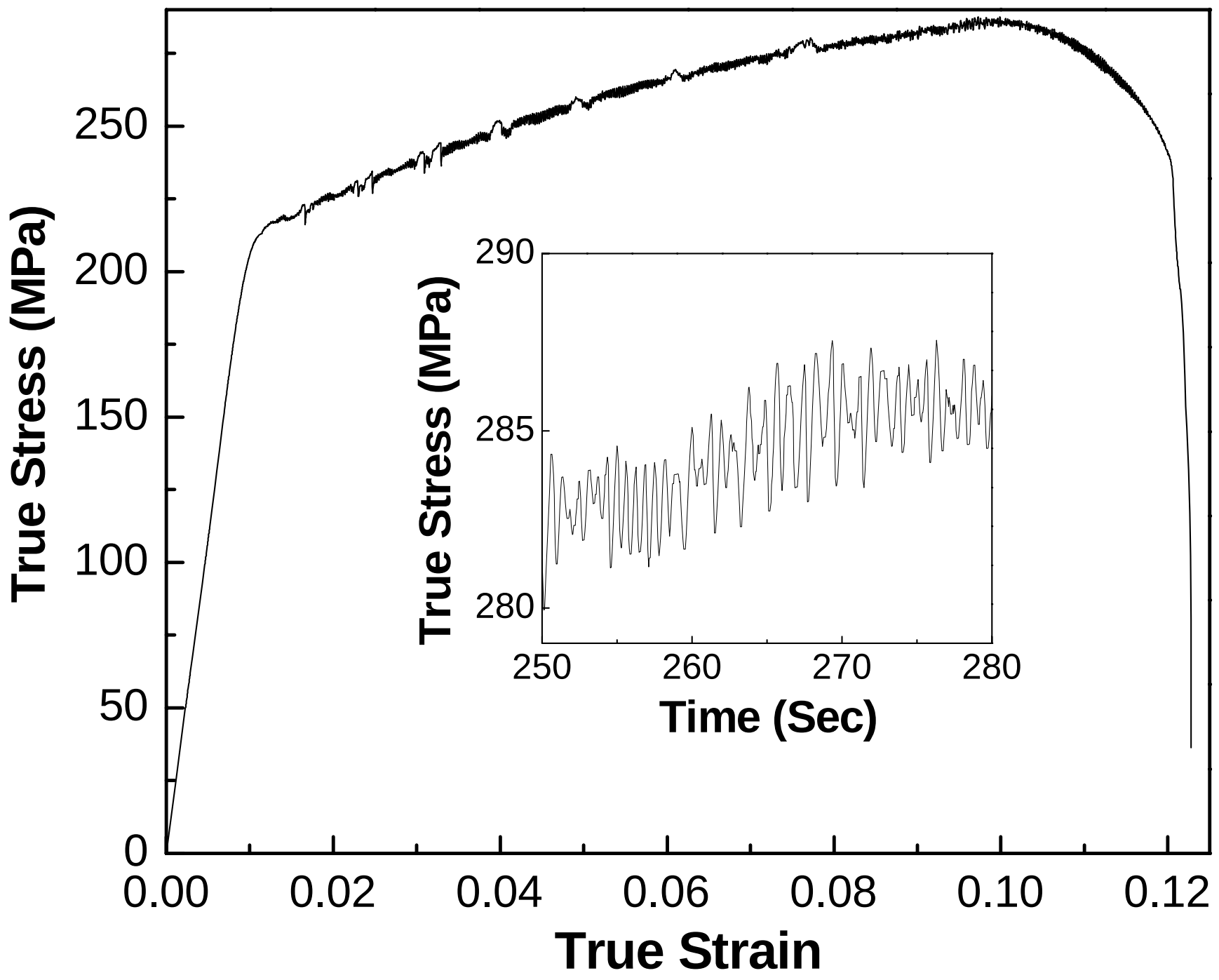

Fig. 1

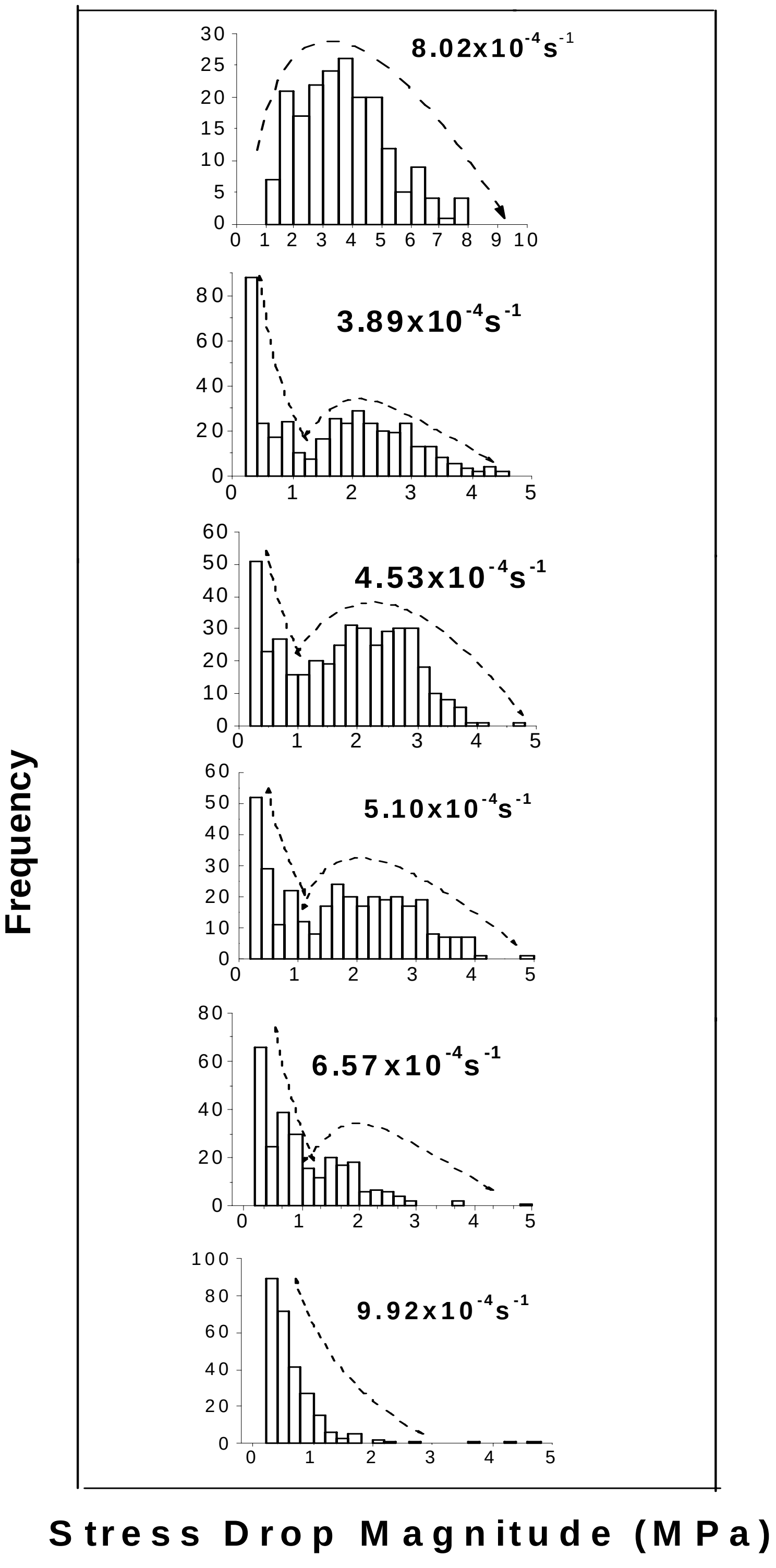

Fig. 2

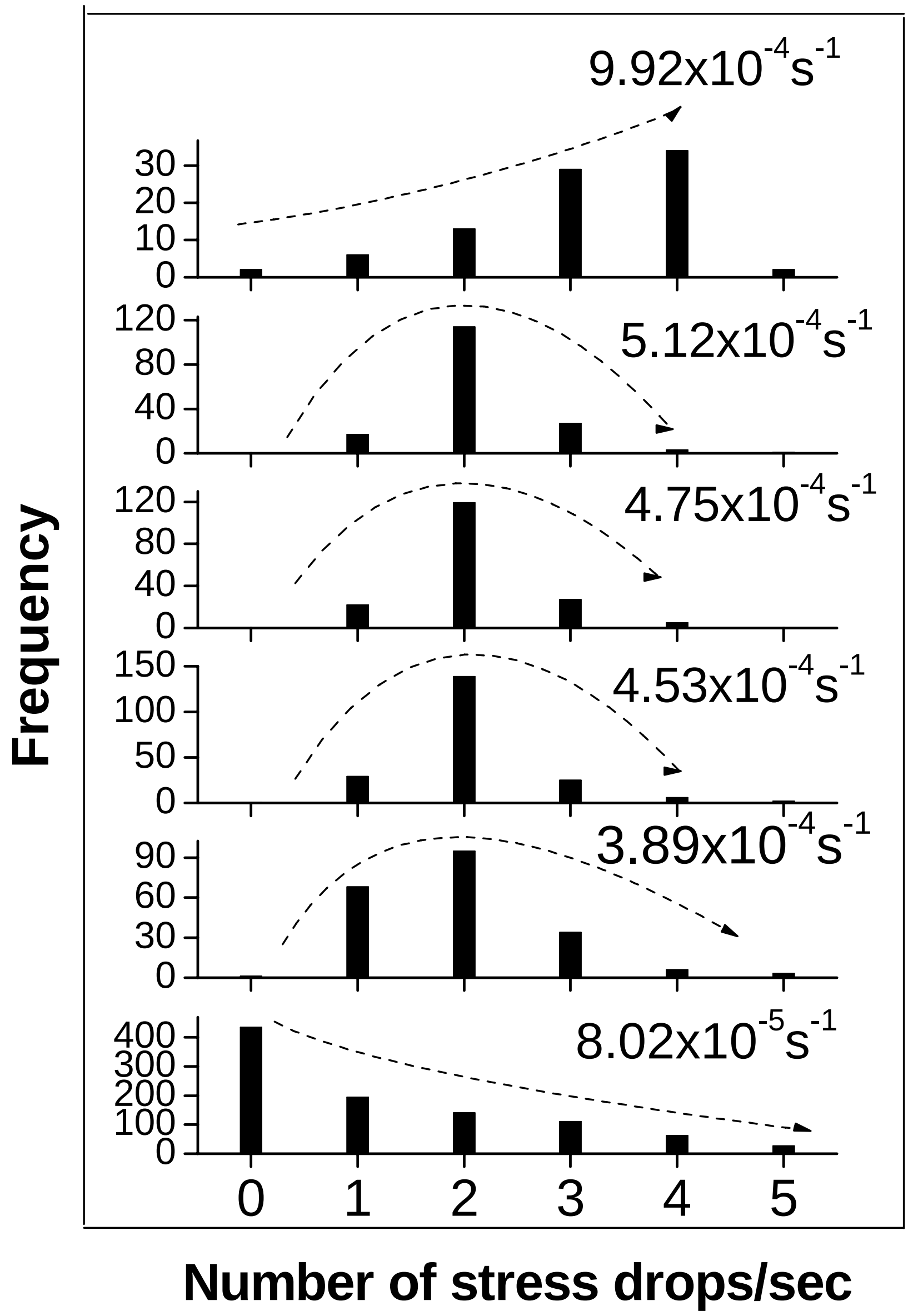
9.92x10^-4s^-1
5.12x10^-4s^-1
4.75x10^-4s^-1
4.53x10^-4s^-1
3.89x10^-4s^-1
8.02x10^-5s^-1
Frequency
0 1 2 3 4 5
Number of stress drops/sec

Fig. 3

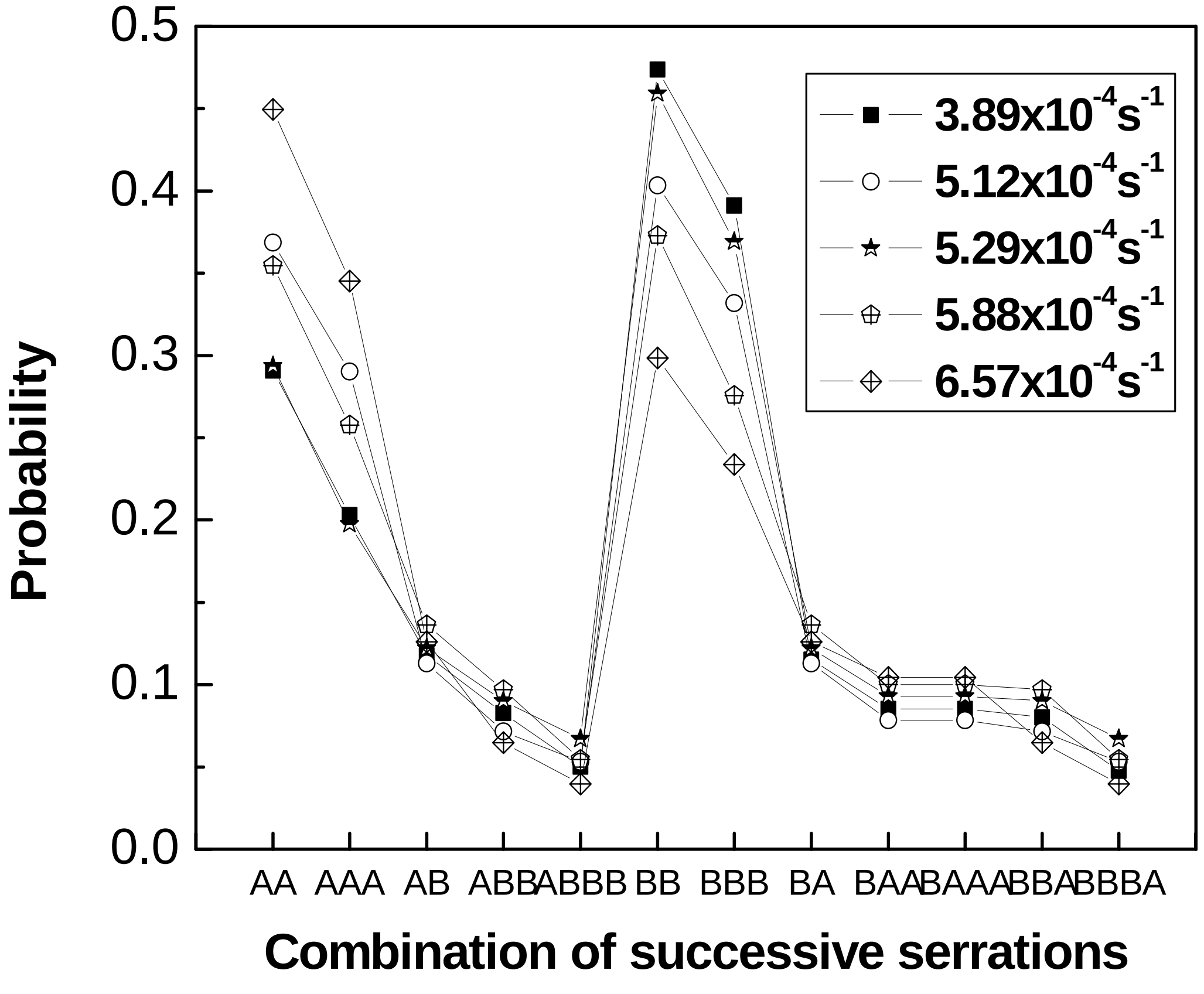
0.5
0.4
0.3
0.2
0.1
0.0
Probability
AA AAA AB ABB ABBB BB BBB BA BAA BAAA BBA BBBA
Combination of successive serrations
3.89x10-4s-1
5.12x10-4s-1
5.29x10-4s-1
5.88x10-4s-1
6.57x10-4s-1

Fig. 4